# Early Pioneers of Telescopic Astronomy in India: G.V.Juggarow and His Observatory

N. Kameswara Rao, A.Vagiswari, & Christina Birdie

*Indian Institute of Astrophysics, Bangalore 560034, India*

**ABSTRACT**

G.V.Juggarow was one of the early pioneers of observational astronomy in India who built his own observatory in 1840 at Vizagapatam (Vizagapatnam). His legacy was continued by his son-in-law A.V.Nursing Row till 1892, his daughter till 1894, Madras Government till 1898, and his grandson till it became inactive in early 1900s. Observations of comets, planetary transits, stellar occultations etc have been continued along with meteorological observations. Celestial photography was also started at the observatory. After 1898 the observatory's activities were re-oriented towards meteorology. The establishment of the observatory, the personality's involved and the final closing of the institution are described here.

## 1 INTRODUCTION

G.V.Juggarow Observatory, established in 1840, was the earliest private modern Indian astronomical observatory that functioned till the turn of the twentieth century. Jesuit priest-astronomers were the real initiators of telescopic astronomy in India, particularly Father J.Richaud who carried out systematic observations of binary stars, comets, zodiacal light, dark clouds etc. at Pondicherry from 1689 (Rao et al 1984)[1]. Although it was claimed that telescope had been used on an earlier occasion by a visiting Englishman, Shakerley, in an unsuccessful attempt to observe transit of Mercury at Surat around 1651 (Kochhar 1989)[2], the credit should go to Jesuits for making a real beginning of telescopic observations. A historical accident led the French Jesuit mathematicians and astronomers to land in India. In 1687 King Louis XIV had sent a team of mathematicians that included a few astronomers like Father Richaud, to Siam on an invitation by King Nair of Siam. A palace coup occurred in Siam that forced the Jesuits to leave the country in a hurry. Three of them,

Father Jean Richaud, Father Jean-Vevant Bouchet and Father Charles Francois Dolu, reached the French colony of Pondicherry (Gopnick 2009)[3]. Father Richaud brought the 12-foot focus telescope he was using in Siam with him to Pondicherry and pursued astronomical observations until his death (Rao et al 1984)[1]. Although introduction of optical telescopes into India, occurred in later half of seventeenth century, establishment of observatories or institutions to pursue observational astronomy did not start until East India Company established its own observatories at Madras and Calcutta more than hundred years later in 1786 and 1825 respectively. Two other observatories were established, through Indian effort, fifty years later, one at Lucknow in 1832 by King Nasiruddin Hayder of Oudh and the other at Trivandrum in 1836 by Raja of Travancore Rama Vurmah. Both these observatories were ostensibly meant for '.. advancement of the noble science's new discoveries as for the defusion of its principles among the inhabitants of India ..' (Nasiruddin Hyder's letter to Lord William Bentinck then Governor General of India -Ansari 1985)[4] and '... that this country should partake with European nations in scientific investigations ..' (Raja Rama Vurmah -Ansari 1985)[4]. However the establishment of these institutions were at the behest of British surveyors (and astronomers) and were headed by British astronomers, Major Wilcox at Lucknow and John Caldecott at Trivandrum. Although several astronomical programmes were initiated by Wilcox at Lucknow, the results could not be published by him and his death in 1848, in a way, resulted in the closing of the observatory (Ansari 1985)[4]. Caldecott also carried out several observational programmes at Trivandrum including observations and orbital computations of comets of 1843 and 1845, observations of Bailey's beads during the solar eclipse of Dec 21, 1843 etc. However bulk of his observations remained unpublished and his death in 1849 resulted in changing the orientation of the observatory from astronomy to meteorology, and astronomy ceased to be practiced there (Ansari 1985)[4]. The two Indian initiatives to pursue observational astronomy did not succeed. G.V.Juggarow observatory on the other hand had a totally different beginning. Juggarow who 'imbibed the keenest interest in astronomy (again a rare accomplishment in men of his position)' (Francis 1915)[5] during his student days at Madras Observatory under the tutelage of T.G.Taylor, then the Government Astronomer, started the observatory at his own expense and own initiative.

## 2 JUGGAROW AND THE BEGININGS OF THE OBSERVATORY

Gode Venkata Juggarow (variously written as Jugga Row, Jagga Row, or Jagga Rao etc.) came from a family of rich and well educated Landlords (*zamindars*). Apparently the family paid more *pesh cash* to the government than any one in the district except the Raja of Vizianagaram. His grandfather, also named Juggarow, was a *dubash* (agent and interpreter) to Mr.John Andrews, Chief at Masulipatam, and he moved to Vizagapatam with Andrews

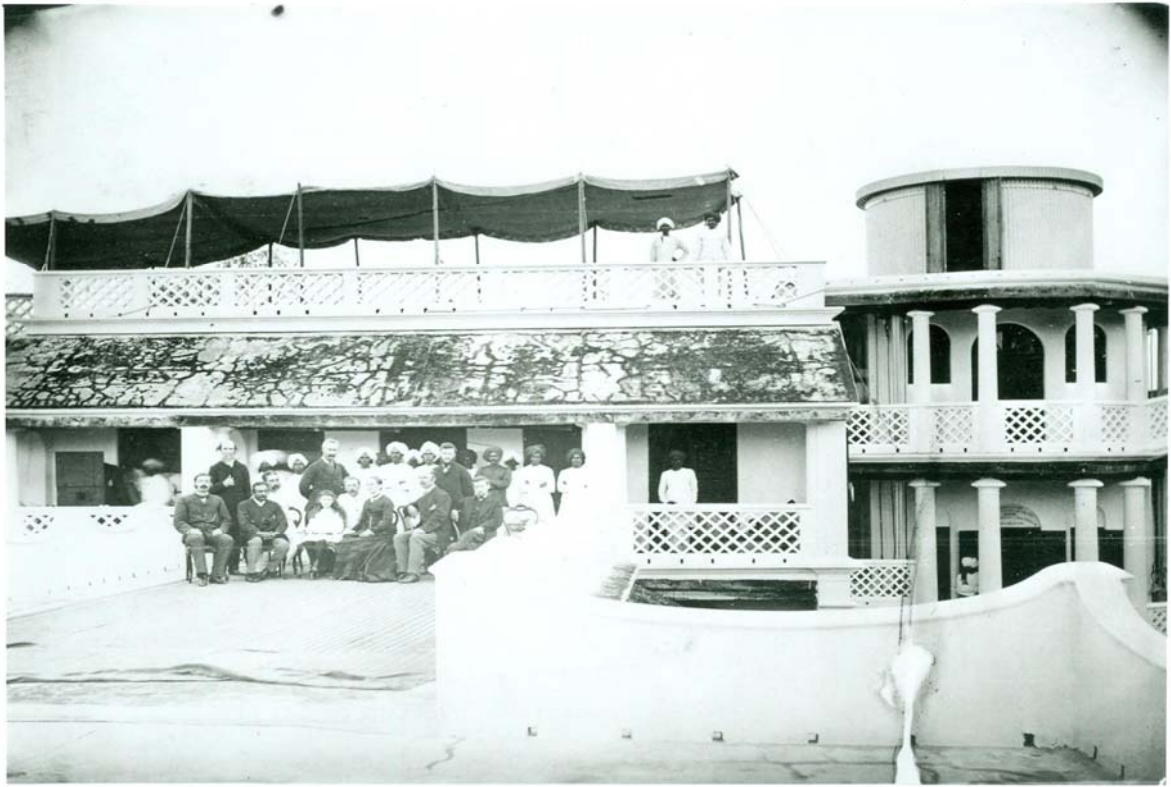

**Figure 1.** The G.V.Juggarow Observatory as seen 1874 on the occasion of transit of Venus. The assembled gathering to witness the event could be seen. A.V. Nursing Row was also present. (Courtesy Royal Astronomical Society)

in 1769. His uncle (father's elder brother) Surya Prakasa Rao was a naturalist and botanist of repute, who could speak and write English language 'uncommonly well' (Francis 1915)[5]. His father Surya Narayana Rao (along with his brother Surya Prakasa Rao) 'are well known to all residents in that neighborhood, for their great intelligence and public spirit, and for the munificence with which they supported the character of native gentlemen' (Morris 1936)[6].

G.V.Juggarow was born in 1817 and was educated at home. Apparently he had a good aptitude for mathematics. During 1833 he seems to have got inspired by astronomy as he says "I selected the divine science of Astronomy as the study and pursuit most congenial to my disposition and best calculated fully to occupy my attention". He left for Madras at the age of 17 years 'from an innate desire of knowledge, he was induced to place himself under the tuition of Mr. Taylor, of the Madras Observatory' (Morris 1936)[6]. By the time he was 19 'we have Mr. Taylor's authority for saying, he is qualified to pass a first class examination in Mathematics in any College in Europe. His knowledge of the English language, also is highly creditable' (Morris 1936)[6]. Training under Taylor at Madras observatory must have provided him an opportunity not only to develop mathematical skills but also to handle astronomical instruments and to conduct astronomical observations. Thomas Glanville Taylor was a well known Government Astronomer at Madras Observatory from 1830 to 1848.

## 2.1 T.G.Taylor

T.G.Taylor grew up among telescopes. His father Thomas Taylor was first assistant at the Royal Observatory, Greenwich, under Astronomer Royal, Maskelyn. His father inducted him into astronomy at the suggestion of John Pond, then the Astronomer Royal, when he was fifteen. Taylor joined the Royal Observatory as a supernumery in 1820 and became a regular staff after two years. Even before his arrival at Madras Taylor acquired a reputation as a good observer. He was selected by Sir Edward Sabine to help him in his pendulum experiments. He assisted Stephen Groombridge and Airy in reductions of circumpolar star catalogues. Sir George Everest was particularly impressed with his skills and required his help in field operations for the Great Trigonometrical Survey. On the recommendation of the Astronomer Royal Pond, East India Company appointed him as Director of Madras Observatory (Fig2). At Madras he got new instruments installed (a five foot transit, a four foot mural circle, and a five foot Dolland equatorial) and initiated regular meridional observations. However, his major achievement was bringing out the famous 'Madras General Catalogue' of 11,015 stars in 1844, which was acclaimed as *'the greatest catalogue of modern times'* by Sir George Airy.

Taylor made several nonmeridional observations, like observations of Comets of 1831, Wilmot's comet of 1845, of Halley's comet in 1836, determination of longitude of Madras, and meteorological and magnetic observations. He also trained many native assistants. Taylor used to provide training for people even to from other institutions like Trivandrum observatory. Taylor was elected as a Fellow of Royal Society on 10 February 1842. He was also a Fellow of Royal Asiatic Society. Apart from publishing his researches in other journals, he was a regular contributor to the Madras Journal of Literature and Science.

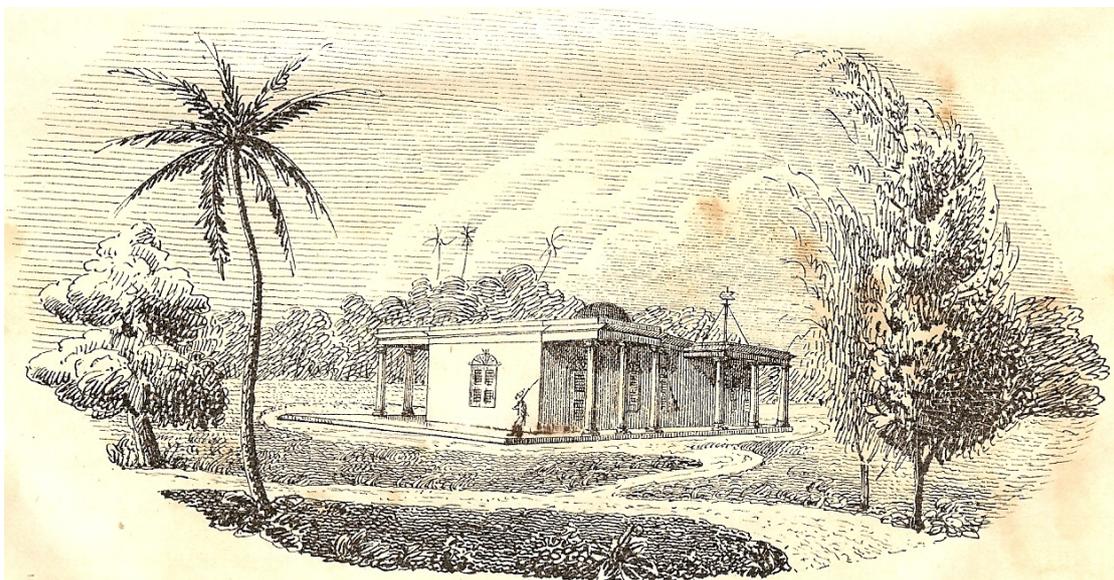

**Figure 2.** Madras Observatory during the period of T.G.Taylor -1830-1848 (IIA Archives).[26]

## 2.2 Juggarow at Madras Observatory

Working with such an accomplished astronomer such as Taylor, it s not surprising that Juggarow got inspired and started his own observatory (Fig1). Taylor and Morris, who was the editor of Madras Journal of Literature and Science, encouraged Juggarow to contribute his researches to the journal. Taylor in his paper on Halley's comet published in MJLS (Taylor 1936)[7] based on observations obtained at Madras observatory acknowledges Juggarow's help 'I take this opportunity of acknowledging my obligations to *Goday Venkata Juggarow*, who rendered me very great assistance in the computations". Juggarow(1936)[8] contributed a note in the same journal titled ' Tables for computing the position of HALLEY's comet' in which he describes his computations of Ephemeris of the comet, from the orbital elements calculated by Taylor from observations obtained at Madras, so as to observe the comet,' when visible to a good telescope if not to the naked eye' at the location of Madras for the month of January 1836 at 10 days intervals. He also provided tables of the mean anomaly corresponding to every 10 minutes of eccentric anomaly as well. It is likely that Juggarow even participated in obtaining the observations.

Starting as early as 1835, Juggarow published several papers on computations of Lunar Occutations, and timings of Moon's rising and setting in the Madras Journal of Literature and Science. These predictions were even verified by observations made at the Madras Observatory[31, 32, 33]. Juggarow(1836)[9] also contributed a paper '*On the Mass of the planet Jupiter*' in which he derives the mass by using the motion of its satellite and Kepler's laws to illustrate in a simpler way to estimate approximate value 302 times earth mass (present value is 317). Morris (1936)[6], the editor of the journal and a FRS, comments patronizingly that he published this paper to illustrate to the readers 'how this great theorem in physical Astronomy is worked out ... further ... being the production of a native of India, among whose brethren intellectual efforts of such character are too uncommon '.

Morris offers considerable praise to young Juggarow as he 'promises to be a distinguished instrument in the good work of elevating the people of India from the character for indifference to intellectual acquirements ... Providence has blessed him with excellent talents, and has placed him in a station of life where his example will have great influence, it is his bounden duty to do all in his power for the promotion of education among his countrymen, with a view to elevating them as a people in the moral and intellectual scale.'

According to Francis (1915)[5] Juggarow was 'on one occasion recommended to act for Mr. Taylor which is a measure of young Juggarow's abilities.

## 2.3 Back at Vizagapatam

After a stay of four years at Madras he returned to Vizagapatam in July 1938, which was commemorated by a present, a statue of Jupiter (probably because he wrote a paper on Jupiter's mass) , which he received from 'Srinivasa Pillai and others, his native friends at the Presidency, who fully appreciating his public spirit evinced by his exertions for the general welfare of the Hindu community, thus express their friendship, esteem and admiration ..' (Vadivelu 1903)[10]. Sriman Srinivasa Pillai was apparently a Dayawant Bahadur. Back at Vizagapatam Juggarow established in 1840 an observatory at his own residence in Daba Gardens (Figure 3) to conduct both astronomical as well as meteorological observations. He also erected the flag staff on the Dolphin's Nose (Figure 5) to provide to the public the time signals. The flag used to be hauled down precisely at 9 A.M to set the time for the station 'for the information of persons who live too far from the fort to hear the report of the gun, and also to afford the shipping in the roads an opportunity of finding the error of their chronometers'(GVJR1896)[11].

He furnished the observatory with a Troughton's Transit Circle and a Chronometer. The Meridian post is at a distance of two miles. He also seemed to have procured optics for a 4.8 inch aperture and 5 feet 8 inch (focus) long telescope from W.S. Jones of London with eyepieces of 40, 60, 80, 120, 200 and 300 power. But he could not get the optics properly mounted into the telescope before his death in 1856 (Nursing Row 1868)[12]. He apparently had great plans for the observatory. The full telescope was later assembled and tested and used by Nursing Row who took over the observatory after Juggarow's death. A photograph of the telescope and its mount is shown in a pamphlet published by Nursing Row in 1868.

Nature of astronomical work carried out at the observatory by Juggarow is not quite clear but he did establish the longitude and latitude of Vizagapatam from his observations. He must have been providing the time corrections from astronomical observations with the transit. However no records are presently available.

Juggarow was also an instrument maker. He invented a splendid Rain gauge (Pluviometer) and was quoted in the 'Buchan's handy book of Meteorology'[13] and also in Chamber's 'Encyclopedia' Vol. VIII, p98-99. The pluviometer was in the form of a funnel having a diameter of 4.697 inches or a receiving area of 17.33 square inches. Since a fluid ounce contains 1.733 cubic inches of water it followed that for every fluid ounce collected by the Gauge the 10th of an inch of rain had fallen. The measure could of course be graduated to any degree of nicety and it may easily be reproduced if required. The cost of the equipment was also very cheap.

He prepared a place near the transit circle for a standard barometer, wet and dry bulb, thermometers, anemometer and an anemoscope constructed by himself on the principles of Hutton's investigations (Nursing Row 1868)[12].

G.V. Juggarow died in 1856 at young age of 39 years leaving his property and the observatory to his daughter Srimathi Ankitham Atchayyamma garu (who became zamindarini of Shermahomededpuram and Yembaram Estates) and son-in-law A.V.Nursing Row.

Nursing Row devoted considerable efforts in promoting astronomy at Daba gardens.

## 3 A.V.NURSING ROW AND THE OBSERVATORY

Ankitham Nursing Row (variously spelt as Nursinga Row, Narasing Row) was born in 1827. His father, Ankitham Sriramulu, was a Dewan to Nawab of Masulipatam, and died soon his son's birth. Nursing Row grew up in Vizagapatam at his maternal grandfather's place and got educated there by private tutors. His teachers were Mr. Porter initially and Rev. J. Hay,D.D later, who apparently also provided him some early training in Astronomy. He joined East India Company quite young and rose to the level of a Deputy Collector in the government. He resigned his position to manage both his (wife's) estates and G.V.Juggarow Observatory. He added a workshop and other amenities to the observatory. The early astronomical observations at the observatory seem to be confined to generation of corrections to the sidereal clock by observing stars with Troughton's Transit Circle as well as providing accurate mean time signals to the Flag staff and Guns on Dolphin's Nose, for civil needs.

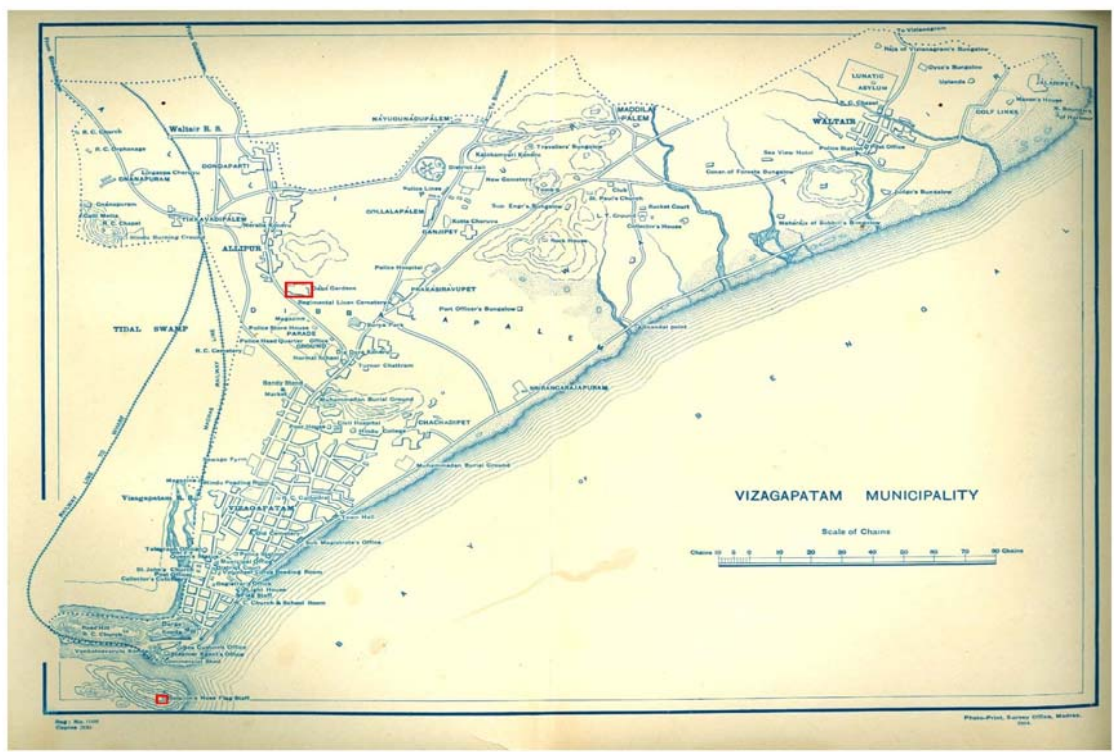

**Figure 3.** The plan of Vizagapatam Municipality as it was in 1914 (Francis 1915)[5]. It shows the locations of Daba Gardens where Juggarow Observatory was situated and the flagstaff on Dolphin's Nose which was used for displaying the time signals.

The first published account of the observatory's scientific activities was described in a pamphlet brought out by Nursing Row on the occasion of the famous solar eclipse of 1868 August 16, which was partial at Vizagapatam. He got a telescope of 4.8 inch aperture, whose optics was earlier procured by G.V.Juggarow, made ready in his own workshop for the occasion. The optics was fitted into the tubes and mounted on 'large and original kind of altazimuth stand' supplied with vernier-circles, levels and slow-motion screw-movements. He also describes that the quality of images are good without chromatic and other aberrations (coma etc.). '*the present telescope will discharge its duties faithfully*'. He goes on to describe his attempts of observing the eclipse. They hoped to see prominences at the limb during the eclipse however they could not see any because of clouds as well as it was only a partial eclipse. He also describes the Indian method (panchang) of calculations of the circumstances of the eclipse at Vizagapatam and also the mythological stories about the eclipses. Figure 4 is the portrait of A.V.Nursing Row.

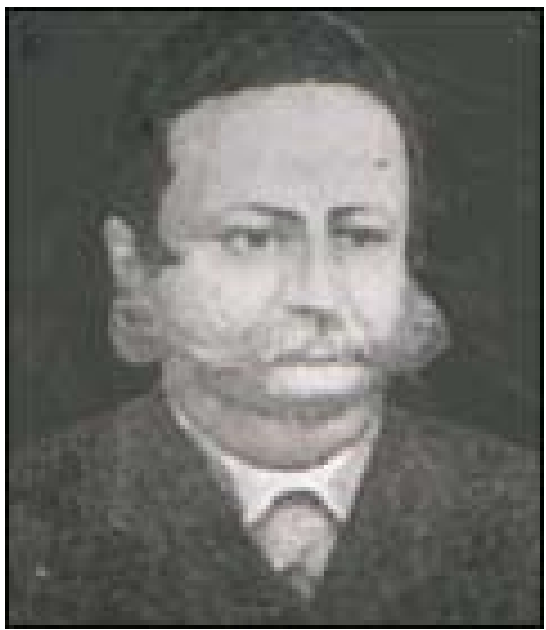

**Figure 4.** M.R.Ry. A.V Nursing Row who essentially equipped, maintained and conducted observations at the observatory until his death in 1892. (Ref.Mrs.A.V.N.College Homepage)[27]

After this beginning, he kept publishing his observations regularly in *Monthly Notices of Royal Astronomical Society* (MNRAS). He kept in regular touch with other British astronomers, most notably, C.Piazzi Smyth of Royal Observatory, Edinburgh and W. Huggins, who communicated several of his observations to MNRAS. On one occasion Piazzi Smyth (1869)[15] writes admiringly about Nursing Row '*a Hindoo gentleman of Vizagapatam, whose family has been much given to science, through two generations. He possesses an extensive Observatory both astronomical and meteorological.*' On another occasion he writes (Piazzi Smyth 1882)[14] '*well-known and liberal-minded, amateur-astronomer, computer, and meteorologist*'.

Nursing Row even in 1868 had plans of adding to the observatory a '6-inch equatorial, with driving clock, micrometer and a *spectroscope*' (Piazzi Smyth 1869)[15]. It is significant that Nursing Row realized the potential of using the spectroscope with a telescope. Unfortunately he never seems to have used a spectroscope.

Soon after the solar eclipse of 18 August 1868, another significant astronomical event, the Transit of Mercury over Sun's disc, occurred on 5th November 1868 which was observed by Nursing Row with his 4.8-inch telescope at Daba Gardens. His account of this event was published in MNRAS (communicated by Piazzi Smyth - Nursing Row 1869)[16]. 'The planet was seen as an intensely dark spot on the Sun's disk'. He shows a diagram illustrating the path of the transit across the disc and gives the timings of the chord. He seems to have observed a curious phenomenon 'when the planet approached half-way of the transit, some of my European friends and myself observed a wavy tint of light darting from the upper edge, disturbed at times, but continued until the planet had passed some distance from the highest point of the line of transit'. It is not clear what caused that wavy phenomenon. He could not see similar behavior during the transit of Mercury on 10 May 1891.

The importance of observations of the transits of inner planets across the Sun's disc is that, by combining observations of the chords of transit trajectories from different geographical locations, one can derive the earth's distance to Sun (the astronomical unit) as proposed first by Edmond Halley in 1716. Realizing the importance of these Nursing Row tried to observe them when ever he could. Apart from 5 November 1868, he made observations of transit of Mercury on two other occasions 7, 8th November 1881 and 10th May 1891 and noted the timings of internal and external contacts (with solar limb) at egress. During such observations in 1881 it is reported 'But at this stage of the proceedings there came pouring into his observatory such a stream of his English lady and gentlemen friends that Mr.Nursing Row confesses he lost his presence of mind as an observer'(Piazzi Smyth 1882)[14]. He also observed the transit of Venus on 9 December 1874 (last contact) which has been discussed in detail by Rathnasree and Kumar (2004)[17]. Incidentally, Samanta Chandrasekhar, a traditional Indian astronomer from Orissa, also calculated by independent methods the timings of the transit of Venus 1874. Both the methods and timings of Chandrasekhar have been assessed by Balachandra Rao and found to be correct (Balachandra Rao 2008)[18].

Partial eclipses of Sun have been monitored at Daba gardens starting 1868 August 16, 1871 December 12 of magnitude 0.629, 1872 June 6 of magnitude 0.972 and 1882 May 17. They were particularly interested in the drop in atmospheric change in temperature, pressure, humidity and the solar radiation. All these were reported in MNRAS through letters to Huggins. He followed his father-in-law's interest in Jupiter by monitoring the morphology of the red spot through his 6-inch equatorial.

Most interesting are observations of Comets by Nursing Row. The spectacular Sun-grazing Great comet of 1882 II, which made history as one of the comets that launched the project of 'carte du ciel' (photographic mapping of the sky)(Gingerich 1992)[19]. A lot has been written about the comet, which is now thought to belong to Kreutz Sungrazers and also called Kamikaze comets (named after the Japanese suicide pilots of World War II). These are the comets that are thought to be plunging into the Sun. The evolution of these comets are studied by Sekanina & Chodas (2007, 2008)[20,21]. Madras observatory observations particularly pre-perihelion are described by (Rao et al 2007)[22]. Nursing Row (1882)[23] presents a colourful description of post-perihelion passage of the comet 'The length of the bright part of tail is between 7° and 8°, above the narrow rays or streaks of light extending more than 12' - concave towards the south and shaped like the *tusk of an elephant*, the thick part being at the greatest altitude'. Interestingly, he says, 'Our Hindoo astronomers predicted the appearance of a comet in the Southern hemisphere …it would possess a bright copper colour like the rising moon, and a long tail. The name given by them to the predicted comet is *Silpacam.* Another comet which was observed at Daba gardens is Comet Pons-Brooks on 1884 January 31.

His efforts of improving facilities at G.V.Juggarow observatory by getting a new Cook equatorial telescope of 6-inch aperture and 7.5 feet focal length that is driven by clock work, and a new 36- inch transit instrument with two setting circles by 'Negretti and Zambra' and a Dent pendulum clock and more importantly communicating his astronomical observations to other astronomers and to Monthly Notices of the Royal Astronomical Society (MNRAS) drew admiration. He was elected as Fellow of Royal Astronomical Society on 4$^{th}$ November 1870. His name was proposed by Astronomer Royal G.B.Airy and W.Huggins, C.Piazzi Smyth and C.Pritchard (RAS Archives)[30]. He was the first Indian to get elected and not Ragoonatha Chary as earlier mentioned by (Rao et al 2009)[24]. He also got elected as a Fellow of Royal Geographical Society the next year 1872. By 1874 before the event of transit of Venus, he got a moveable dome of 12 feet diameter and 9 feet high installed. It is quite common for Vizagapatam to get hit by cyclones. The observatory used to record the rain fall, the wind velocities etc. One of the worst such cyclones occurred on 1876 October 7th evening lasting the whole night. The winds were so strong that the new dome of the observatory -a corrugated iron structure which had been placed in position but not riveted down was blown away to a distance of 33 feet.

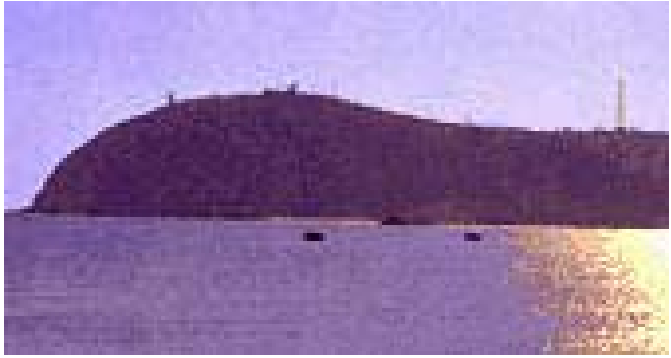
**Figure 5.** Dolphin's Nose where the flagstaff and time-gun used for time signals are located.[28]

Nursing Row was also a very public spirited person. When the time-gun firing from Dolphin's Nose was stopped in 1871 by the government, he took the initiative of maintaining it by paying the expenses from his private funds. He also got a new and expensive Flagstaff at Dolphin's Nose in 1886(See Fig5). The notification in the District Gazette dated 1st May 1886 says 'It is hereby notified for the information of the public that Mr. A.V. Nursing Row has erected an expensive and durable Flagstaff on the Dolphin's Nose and the time signals are now hoisted on it under his orders every morning between 8 and 9 o'clock to indicate time....' He also regularly furnished the Meteorological Reporter with daily reports of meteorological conditions. Because of such activities he was honoured by the Government with the title of Rai Bahudur'. He was also appointed as Honorary Meteorological Reporter to the Government of India for Vizagapatam. He donated a sum of 1,15,000 Rupees as an endowment to the then Hindu college when it was in financial difficulties. The college was later named as Mrs. A.V.N. College.

His enthusiasm for astronomy was such that he was always trying to improve the facilities at the observatory. Recognizing the benefits of photography to astronomical observations he 'commenced making arrangements to place the observatory on a permanent basis as well as to erect a Celestial Photographic Observatory, but he died before completing this work (GVJGR 1896)[11]. Nursing Row died on 18 June 1892.

## 4 TRANSITION PERIOD AND A.V.JUGGA ROW

In 1884 Mrs Nursing Row proposed to endow the institution with a fund of three lakhs and hand it over to trustees under the control of the Madras Government. However government was unable to accept the position at that time. After the death of Nursing Row, his wife Sri Ankitham Atchayyamma Garu took charge of the observatory and conducted with great 'liberality'. She brought out the 'Results of Meteorological Observations 1894' under the auspicious of G.V.Juggarow Observatory 'maintained by Mrs. A.A.Nursing Row, zamindarini of Shermahamudpuram and Yembaram Estates'. She continued to supply meteorological results to the Government of India, Meteorological Reporters at Bengal and Madras, and, in addition, erected a Celestial Photographic Observatory with a photographic telescope. In accordance with the wishes of her father as well as her husband, she handed over the Observatory and the Dolphin's Nose Flagstaff on the 8th November, 1894 to the Government of India with an endowment of 3 lakhs of rupees for the permanent maintenance of the institution. The management of the institution was to be done by a committee comprising the Collector for the time being, the Meteorological Reporters of the Governments of Bengal and Madras, the Government Astronomer and others. Soon after it appears that the committee appointed Mr.Bion as Astronomer for a period of three years to run the observatory. The committee till 1898 consisted W.O.Horne as chair and J.Elliot M.A.,F.R.S.,C.I.E., as Secretary.

The committee (Government) also started a new Observatory at Waltair to conduct meteorological observations. The routine observations which consisted of measurement of double stars, celestial photography, and daily time service continued. By now a telegraphic line was established for the time-gun firing from the observatory to the Dolphin's Nose. The error of the sidereal clock being determined by transits of clock stars and the flash of the time gun and its error being observed every night from the observatory.

The binary stars being observed (we assume the angular separations and position angles for orbital measurements) in 1897 were 41 Aqr, $\mu$ Cygni, $ry$ Ari, $ry$ Cet and 326 Ari.

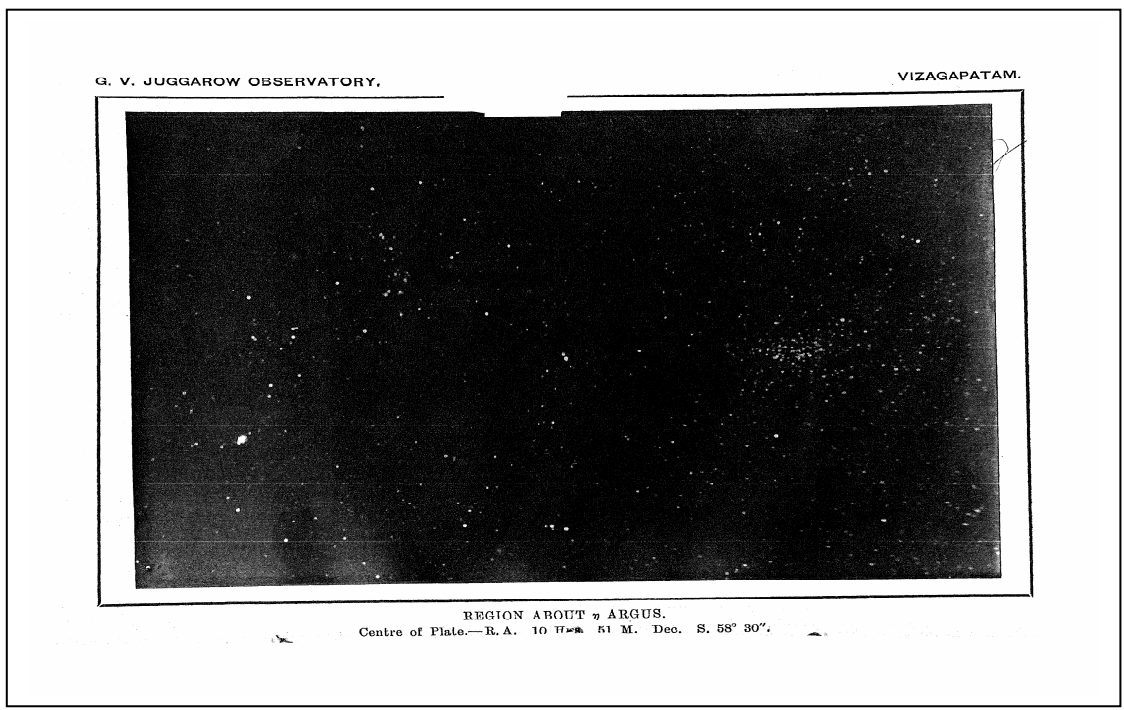

**Figure 6.** The photograph obtained at G.V.Juggarow observatory of 77 Carina region.(GVJR 1899)[29]

## 4.1 Photography of night sky objects

G.V.Juggarow observatory must have been the first one in India to systematically obtain photographic observations of the night sky. Photographs of various clusters and nebulae were taken with the small photographic telescope, the best being perhaps that of the region near 71 Argus (now $_q$ Caraina) (Figure 6). 'As the lens of the camera is small one photograph of nebulae are not as successful as those of clusters' (the focal length of the telescope might have been large too!). Prof. H.H.Turner was helping and advising in the selection of various kinds of plates which were being tested at the observatory. Apparently a sidereal-time watch was ordered for the photographic observatory. The camera and chronometer was kindly ordered for the observatory by Mr. Michie Smith, Government Astronomer at Madras but they did not arrive by the end of the year. Contact with Madras observatory was restarted after G.V. Juggarow took over.

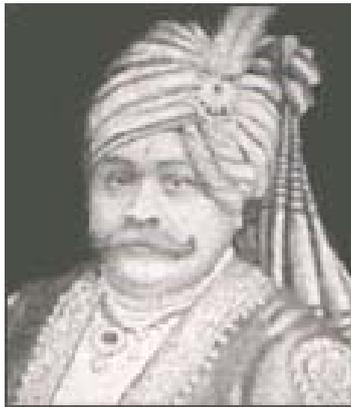

**Figure 7.** Sri A.V.Jugga Row Bahadur grandson of the founder. He was Vice-President of Astronomical Society of India for 1911- 1912.[27]

## 5 A.V.JUGGA ROW AND FADING OF THE OBSERVATORY

Subsequently M.R.Ry. A.V.Jugga Row regained the control of the observatory and its endowments.

Sri Rajaa A.V.Jugga Row Bahadur Garu was born on 4th February 1866 at Vizagapatam (Vadivelu 1903)[10]. His early education was done at London Mission High school in Vizagapatam and at home by his father. He had special training in astronomy and meteorology. The partial solar eclipse observations of 1882 May 10 were done by A.V.Jugga Row and Verabadroodoo (Nursing Row 1882)[25]. He succeeded to the management of the estate in October 1898. Soon after this he took over the observatory and started maintaining it. He was more interested in Meteorological Observations. 'He opened a magnetical observatory', one of the three in the country. 'He has also opened a Seismological Observatory, which is very rare institution in any country for measuring the current of earthquakes'.

Mr. A.V.Jugga Row was elected Fellow of Royal Astronomical Society on 8$^{th}$ May 1900, His name was proposed by E.V.Struobel, W.Huggins and H.H. Turner (RAS Archives)[30]. He became a number of the Royal Meteorological Society, Royal Colonial Institute and Society of Arts in 1900 when he was visiting England and Queen Victoria. During this tour he apparently visited various observatories and studied their working.

He was also elected as one of the Vice-Presidents of Astronomical Society of India from 1911 to 1912.

The astronomical activity slowly (**or abruptly?**) faded away at Daba Gardens during A.V.Jugga Row's time. It is not clear (no records) what lead for the Government to relinquish the management of the observatory.

Presently, Dolphin Hotel, a three star hotel, exists at Daba Gardens where once G.V.Juggarow Observatory was (stood).

## 6 CONCLUDING REMARKS

G.V.Juggarow Observatory is an unique institution that was created by an inspired individual, Juggarow, who got fascinated by the allure of the night sky and acquired the taste of making observations of celestial sources by working with an accomplished professional astronomer. He, in turn, seems to have conveyed this excitement of rediscovery of the nature and splendor of celestial sources, when rest of world had gone to slumber, to his son-in-law, Nursing Row who continued to carry this fascination of working with the telescopes till his death. It was really Nursing Row who provided and developed not only the facilities at the observatory but also conveyed his enthusiasm of whatever he has done to other astronomers and to journals irrespective of whether it was a simple phenomenon of partial solar eclipse or looking at the red spot on Jupiter or helping to measure the astronomical unit by tracking the transit of planets over the Sun's disk. Nursing Row was also very aware and sensitive to the contemporary developments in astronomy. He planned to acquire a spectroscope realizing the importance of it in physical astronomy. He made efforts to start photographic observations of the sky, which could be materialized only after his death. Juggarow observatory was the only observatory in India which started systematic photographic observations of the night sky (star clusters and nebulae). Unfortunately, it did not continue.

Contrary to the general belief among science historians that 19th century observational astronomy simply failed to take off under Indian auspices, as though Indians had some inhibition to do observational astronomy, it did flourish in some quarters. It is often forgotten that observational astronomy needs a lot of resources, not only human initiative and ideas, but great deal of technical support to put them into practice. To pursue photographic observations one needed continuous supply of sensitive photographic plates, chemicals, apart from telescopes with cameras and plate holders, dark rooms etc. which had to be imported and properly preserved (so that emulsions do not get damaged). It was a tough job. It is remarkable that G.V.Juggarow observatory functioned with enthusiasm for about sixty years, when two other contemporary Indian sponsored observatories at Lucknow and Trivandrum barely managed to survive for ten years and that too without being able to publish any results.

It is not clear, presently, how the observatories management changed from the Madras government to A.V.Jugga Row Garu and more importantly how and why the observational programmes enumerated in the 1898 annual report got discontinued. It was a pioneering effort by G.V.Juggarow and A.V.Nursing Row to demonstrate Indian effort at modern observational astronomy.

# 7 ACKNOWLEDGEMENTS

We acknowledge with thanks Tamil Nadu Archives for the 1868 report of Nursing Row and Royal Astronomical Society for allowing us to use the 1874 photograph of Daba Garden observatory. We are thankful to Peter Hingley of RAS for sending us the picture and information about Nursing Row's election to FRAS.. This research has made use of the SIMBAD database, operated at CDS, Strasbourg, France.

Our sincere thanks to our colleagues Drs. Ramesh Kapoor, A.V.Raveendran, A.B. Varghese and Mr. P.Prabahar for the help and encouragement.

Finally we would like to thank Department of Science and Technology, Government of India for their financial support through the project SR/S2/HEP-26/06.

Dr.N.Kameswara Rao
Indian Institute of Astrophysics,
Koramangala, Bangalore – 560 034, India
*For correspondence.
e-mail: nkrao@iiap.res.in